# Comment on a paper by Lapaire and Sipe

Morton H. Rubin

*Department of Physics, University of Maryland, Baltimore County, Baltimore, Maryland 21250, USA*

In a recent paper [quant-ph/0607008] Lapaire and Sipe argue that one can discuss interference experiments using entangled photons in terms of single photon wave functions. Furthermore, they argue that contrary to the claim of the authors of the postponed compensation experiment [2], the single photon wave functions overlap on the beam splitter when interference is observed. In this comment, we show that the claim in [2] is correct and we argue that the idea of single photon wave functions in entangled states is misleading.

A bipartite entangled system in a pure state can be represented as a superposition of single particle pure states using, for example, the Schmidt decomposition,

$$\Psi = \sum_j c_j |\phi_j\rangle |\chi_j\rangle. \qquad (1)$$

This representation is not unique; however, if I understand the philosophy of [1], the authors wish to ascribe single particle state vectors to each particle. Of course, it is possible to make definitions at will, but I believe that their definition is conceptually misleading except in the case of separable states. I prefer to follow the approach of Schrödinger. In the classic paper in which he introduced the term entanglement [3], Schrödinger defines the state vector (which he refers to as the $\psi$-function) of a system as an encoding of the maximum total knowledge describing the state of a system. He argues that for subsystems of a system in an entangled state no such function exists. Schrödinger points out that if one is limited to conditional statements, for example, from Eq. (1) *if* subsystem two is in the state $|\chi_1\rangle$, *then* subsystem one is in the state $|\phi_1\rangle$, the notion of a state vector containing maximal knowledge of subsystem one ceases to apply. This is the essence of entanglement. In fact, the only single particle state that can be uniquely ascribed to subsystem one is that obtained by tracing away the state of two,

$$\rho_1 = tr_2 |\Psi\rangle\langle\Psi|. \qquad (2)$$

This state is unique in that it is invariant under all unitary transformations applied to particle 2. This corresponds to only measuring the subsystem one and discarding two. With this state, properties of subsystem one may be uniquely described at the cost of losing all the information about entanglement. In general, even the conditional statement about the state of particle 1 given the state of particle 2 lacks precision unless the state vectors $\{|\chi_j\rangle, j = 1, 2, \ldots\}$ are an orthogonal set as in the Schmidt representation. Finally, wave functions or probability amplitudes are generally complex functions derived by expressing state vectors in a convenient orthonormal bases. Such representations do not change the essence of the above discussion.

Let us now turn to the specific problem discussed in [1], namely, trying to ascribe a wave function to photons. As the authors point out, it is well known that you cannot do this; however, it is very useful to have a space-time picture of experiments involving relativistic quantum fields. From the beginning of quantum optics it has been found useful to use correlation functions to provide such a picture. In particular, for spontaneous parametric down conversion (SPDC), such a correlation function is

$$G^{(2)} = \langle \Psi | E_2^{(-)} E_1^{(-)} E_1^{(+)} E_2^{(+)} | \Psi \rangle = \left| \langle 0 | E_1^{(+)} E_2^{(+)} | \Psi \rangle \right|^2 \tag{3}$$

where $|\Psi\rangle$ is the two photon state vector for some particular SPDC source, $|0\rangle$ is the vacuum state, and $E_1^{(+)}$ is the positive frequency electric field operator at some point of space-time. The fields are written in the Heisenberg representation and satisfy the Heisenberg equations of motion, which are just Maxwell's equations. These operators are determined by the boundary conditions imposed by the particular experiment being analyzed. The quantity $A_{12} = \langle 0 | E_1^{(+)} E_2^{(+)} | \Psi \rangle$ was called the biphoton by Klyshko [4] and is also called the two-photon amplitude. It contains the maximum information about the results of measuring two photon correlations for $\Psi$. The biphoton can be sketched under some simple conditions [5]. In interferometric experiments, the biphoton wave function is generally a superposition $A_{12}^{(a)} + A_{12}^{(b)}$ and everyone agrees that to see interference effects these two amplitudes must overlap, that is, $A_{12}^{(a)*} A_{12}^{(b)} \neq 0$. In general there are two space-time points involved so that this interference is a non-local effect. The classic experiment for this is the HOM interferometer experiment [6]. There are many ways to reduce the two photon state to a single photon state, indeed, this is the basis of much work in defining heralded single photons [7]. In analogy with the discussion above, the only unique correlation function for the single photon 1 is

$$G^{(1)} = tr E_1^{(-)} E_1^{(+)} \rho_1$$
$$\rho_1 = \sum_{\vec{k},\lambda} \langle \vec{k},\lambda | \Psi \rangle \langle \Psi | \vec{k},\lambda \rangle \tag{4}$$

where the summation is over a complete set of orthonormal states for photon two. It is worth noting that this is different than simply integrating Eq. (3) over all possible positions of detector two. The difference is the appearance of the intensity in the integrated version of Eq. (3). For quasimonochromatic waves, this difference is minor, but in general it is not.

Consider now the specific example of postponed compensation. We refer the reader to either [1], [2], and, especially, [8] for details. We discuss the simplest case of a Gaussian pump of bandwidth $\sigma_p$ centered at angular frequency $\omega_p$ generating the SPDC into two directions so we can make the problem essentially one-dimensional. For this case, the biphoton can be written as

$$A_{12} = v(t_+) u(t_-)$$
$$v(t) = A e^{-\sigma_p^2 t^2 / 2} e^{-i\omega_p t} \tag{5}$$

where A is a constant, $t_+ = (t_1 + t_2)/2$ and $t_- = t_1 - t_2$ with $t_j$ denoting the retarded time at a point in the jth beam, that is, if a photon is detected at the time $\tau_j$ at the optical path length $r_j$ from the source, then $t_j = \tau_j - r_j/c$. It is assumed that the signal and idler beams are degenerate. The function $u(t)$ describes the details of the source, the only property of it that we need is that it has finite support so that it is non-zero only when $0 < t < \tau_0$. We have ignored polarization although it plays an important role in the experiment, it is not important for the issues discussed here. For the postponed compensation experiment if we were to place our idealized point detectors after the beam splitter but before carrying out the compensation, we would find that the coincidence counting rate is given by

$$R_c = R_0 e^{-\sigma_p^2(t_+ + \tau_1/2)} |u(t_- + \tau_1) - u(-t_- + \tau_1)|^2 \qquad (6)$$

in order for there to be interference we require $0 < \tau_1 < t_0$. If $\tau_1 > t_0$ the two biphoton amplitudes do not overlap after the beam splitter, implying that they leave the beam splitter separately. Also, if one uses the definition of the single photon state defined above, that is by $G^{(1)}$, then it is easy to show that the signal and idler photons arrive at the beam splitter separately. This was what many people did not believe, when the experiment was carried out. The remainder of the discussion to restore the overlap is standard and will not be repeated.

Let us now repeat the analysis, using notation similar to that in [1]. The biphoton can be written as

$$A_{12} = A \int_{-\infty}^{\infty} dt\, e^{-\sigma_p^2 t^2/2} e^{-i\omega_p t} V_s(t_1 - t) V_i(t_2 - t) \qquad (7)$$

where $V$ is the function defined to be the single photon wave amplitude, and s and i refer to signal and idler, respectively. Some care is required in this interpretation since there is not a single wave function but a continuum of them parameterized by t. Furthermore, these are not orthogonal and so at the outset do not even lend themselves to the "*if-then*" conditional interpretation. After the beam splitter the relevant part of the amplitude becomes

$$\begin{aligned} A_{12} &= A \int_{-\infty}^{\infty} dt\, e^{-\sigma_p^2 t^2/2} e^{-i\omega_p t} \left( V_s(t_1 - t + \tau_1) V_i(t_2 - t) - V_s(t_2 - t + \tau_1) V_i(t_1 - t) \right) \\ &= A \int_{-\infty}^{\infty} dt'\, e^{-\sigma_p^2 (t_+ - t')^2/2} e^{-i\omega_p (t_+ - t')} \left( V_s(t' + t_-/2 + \tau_1) V_i(t' - t_-/2) \right. \\ &\quad \left. - V_s(t' - t_-/2 + \tau_1) V_i(t' + t_-/2) \right) \end{aligned} \qquad (8)$$

where in the second line we have introduced $t_\pm$, and changed the integration variable. In Fig. 1 we illustrate where the two terms in the integrand are non-zero assuming that $V(t) \neq 0$ only for $0 < t < t_0$. The solid lines show the regions where the $V_i$ are non-vanishing, 1 refers to the first term in Eq. (8) and 2 to the second. The dashed lines show where the $V_s$ are non-zero when there is a delay $\tau_1$ in one arm of the HOM interferometer. We see that for a finite coincidence window and large enough $\tau_1$ that the signal and idler terms do not overlap at detectors placed after the beam splitter. It is again clear that this can be compensated for, as the authors of [1] acknowledge. Although it is common practice to draw figures like Fig. 2 of [1], they are crude representation of entangled photons that are very misleading, especially to non-experts. There are no little wave functions in each arm.

Finally, I do not understand why the authors of [1] state that the Feynman path technique is not rigorous. The idea of distinguishable and indistinguishable paths can be made quite rigorous and is quite edifying [9]. It is especially useful when studying quantum imaging with SPDC sources.

It is a curious feature of quantum theory that everyone agrees how to make correct calculations while maintaining different pictures of the content of the calculations. Aside from my disagreement with the authors of [1] over their statement about the overlap issue in the experiment in [2], the main difference I have with them is one of interpretation. Furthermore, everyone agrees that one can define mathematical single particle wave amplitudes for entangled states, but the issue is whether one can ascribe any kind of physical reality or explanatory power to them. I believe that one should hew as close as possible to Schrödinger's remarkable insight about entanglement and avoid any picture that would mislead the unwary; it is for this reason that the authors of [2] concluded that the postponed compensation experiment can not be described in terms of single photon wave functions.

Acknowledgments: I wish to thank A. Garucio, Y-H Kim, T. B. Pittman, G. Scarcelli, A. Sergienko, D. V. Strekalov, and Y. H. Shih for their comments on this note.

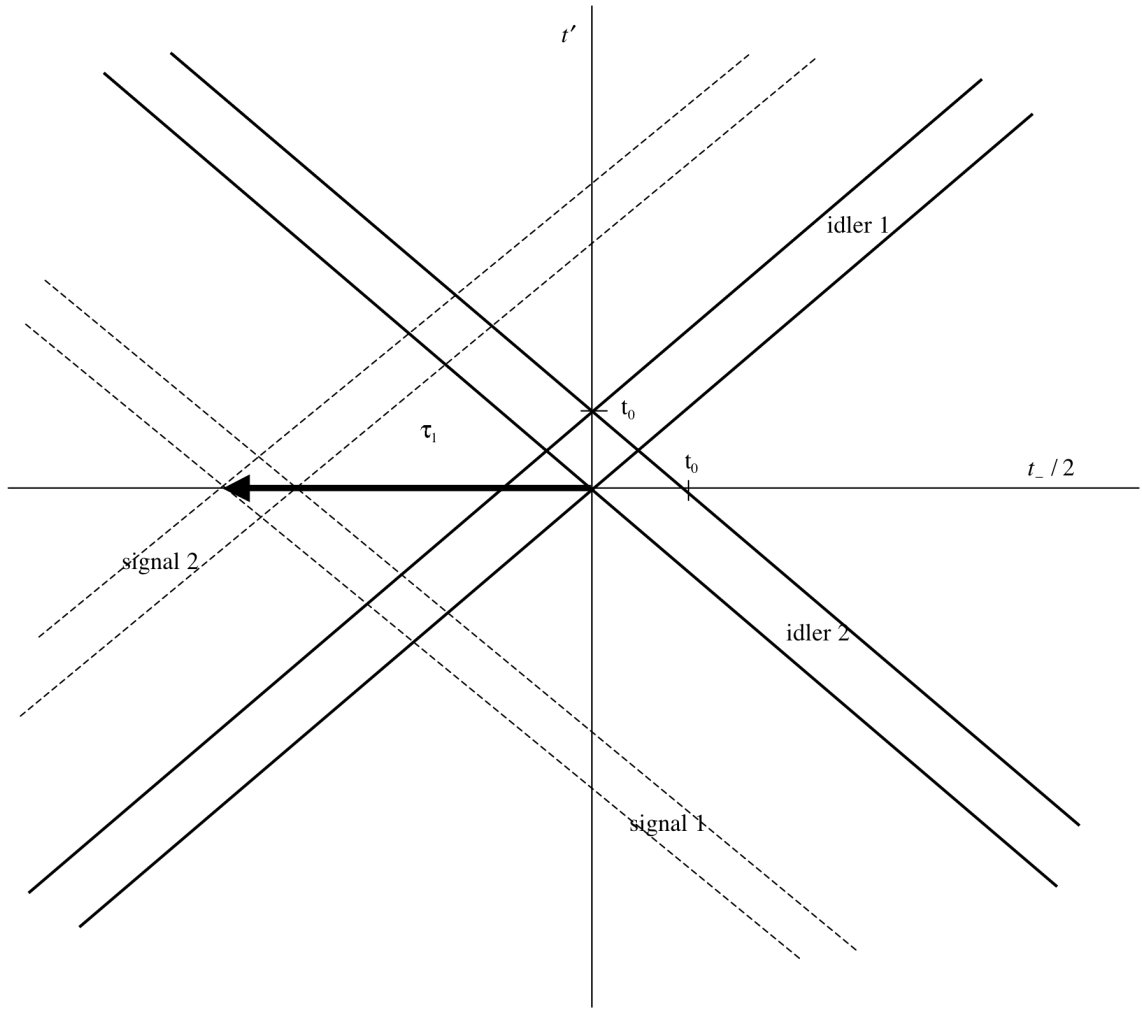

Figure 1: The two sets of parallel solid lines represent the regions of retarded time where the idler amplitudes $V_i$ are non-zero in the biphoton wave packet. When there is no delay, $\tau_1 = 0$, the signal amplitudes $V_s$ overlap the idler wave amplitudes with signal 1 (2) and idler 2 (1) having the same support . When a delay is introduced, $\tau_1 > 0$, in the signal arm of the output of the SPDC source, the support region of the signals is displaced into the regions between the dashed lines.